\documentclass[journal,onecolumn]{IEEEtran}

\usepackage{epsfig}
\usepackage{graphicx}
\usepackage{graphics}
\usepackage{epstopdf} 
\usepackage{pgf, tikz}
\usepackage{listings}
\usetikzlibrary{calc,shapes.geometric}
\usetikzlibrary{snakes,backgrounds}
\usetikzlibrary{quotes,angles}
\usetikzlibrary{automata, positioning}
\usetikzlibrary{shapes,decorations,arrows,calc,arrows.meta,fit,positioning}
\tikzset{
    -Latex,auto,node distance =1 cm and 1 cm,semithick,
    state/.style ={ellipse, draw, minimum width = 0.7 cm},
    point/.style = {circle, draw, inner sep=0.04cm,fill,node contents={}},
    bidirected/.style={Latex-Latex,dashed},
    el/.style = {inner sep=2pt, align=left, sloped}
}
\usepackage{cite}
\usepackage{amssymb}
\usepackage{multirow}
\usepackage{tcolorbox}
\tcbset{colframe=black,colback=white,colupper=black,
fonttitle=\bfseries,nobeforeafter,center title,size=small,arc=0pt,outer arc=0pt}

\ifCLASSINFOpdf
\else
\fi

\usepackage[cmex10]{amsmath}

\def\clock{\n=\time \divide\n 60
  \m=-\n \multiply\m 60 \advance\m \time
  \ifnum \n>12 \advance\n -12 \fi
   \number\n.\twodigits\m~\ampm\time}
\def\ampm#1{\ifnum #1< 720 am\else pm\fi}
\def\twodigits#1{\ifnum #1<10 0\fi \number#1}

\usepackage{caption}
\usepackage{subcaption}

\def\nexto{\kern -0.54em}
\def\mean{{\rm {I\ \nexto E}}}
\def\prob{{\rm {I\ \nexto P}}}

\def\dist{\stackrel{d}{=}}

\def\pfa{{\rm P_{FA}}}
\def\pd{{\rm P_{D}}}

\def\P{{\cal P}}
\def\ray{{\rm Ray}}

\newcount\m \newcount\n
\def\clock{\n=\time \divide\n 60
  \m=-\n \multiply\m 60 \advance\m \time
  \ifnum \n>12 \advance\n -12 \fi
   \number\n.\twodigits\m~\ampm\time}
\def\ampm#1{\ifnum #1< 720 am\else pm\fi}
\def\twodigits#1{\ifnum #1<10 0\fi \number#1}

\definecolor{codegreen}{rgb}{0,0.6,0}
\definecolor{codegray}{rgb}{0.5,0.5,0.5}
\definecolor{codepurple}{rgb}{0.58,0,0.82}
\definecolor{backcolour}{rgb}{0.95,0.95,0.92}
 
\lstdefinestyle{mystyle}{
    backgroundcolor=\color{backcolour},   
    commentstyle=\color{codegreen},
    keywordstyle=\color{magenta},
    numberstyle=\tiny\color{codegray},
    stringstyle=\color{codepurple},
    basicstyle=\footnotesize,
    breakatwhitespace=false,         
    breaklines=true,                 
    captionpos=b,                    
    keepspaces=true,                 
    numbers=left,                    
    numbersep=5pt,                  
    showspaces=false,                
    showstringspaces=false,
    showtabs=false,                  
    tabsize=2
}
 
\begin{document}


\title{Passive Sonar Sensor Placement for Undersea Surveillance}
\author{Graham  V. Weinberg and Martijn van der Merwe}
\maketitle

\markboth{Passive Sonar Sensor Placement: \today}%
{}

\begin{abstract}
Detection of undersea threats is a complex problem of considerable importance for maritime regional surveillance and security. Multistatic sonar systems can provide a means to monitor for underwater threats, where fixed sensors, towed arrays and dipping sonars may be utilised for this purpose. However, it is advantageous to deploy passive sensors to provide a stealthy early warning system. Hence this paper is concerned with determining where a series of passive sonar sensors should be situated in order to provide an initial threat detection capability.
In order to facilitate this it is necessary to derive a suitable expression for the probability of threat detection from a passive sensor. This is based upon considerations of the passive sonar equation. It will be demonstrated how the stochastic aspects of this equation may be modelled through appropriate random variables capturing the uncertainty in noise levels. Subsequently this is utilised to produce the system-level probability of threat detection. Since the threat location is also unknown an appropriate statistical model is introduced to account for this uncertainty. This then permits the specification of the probability of detection as a function of sensor locations. Consequently it is then possible to determine optimal sensor placement to maximise the threat detection probability. This provides a new way in which to determine whether a surveillance region is covered adequately by sensors. The methodology will be illustrated through a series of examples utilising passive sonar characteristics sourced from the open literature.
\end{abstract}

\section{Introduction}
\label{sec:intro}
The primary objective of this paper is to develop an algorithm to determine the optimal placement of passive sonar sensors which are being deployed to provide an undersea early warning system against threats such as submarines. The utility in passive sensors is that they do not ping so are not detectable by a potential threat. However they can be severely affected by ambient noise, making detection of low-signature threats a challenging exercise. This means that the determination of the best location of passive sensors is important from an operational perspective. A passive sonar system may be used to establish initial threat detection and bearing, information which can then be utilised by systems deploying active sonars to facilitate localisation and tracking, as well as threat engagement \cite{fillinger10}.

Detection, tracking and defeat of submarines is known as anti-submarine warfare (ASW) which continues to be a topic of significant importance, especially given the fact that nuclear-powered submarines are significantly more difficult to detect than their conventional counterparts \cite{tyler92, miasnikov94}. 
An early investigation of ASW is \cite{danskin68} who developed a game-theoretic approach to the detection problem. A dipping sonar is often used as a first stage in ASW operations and as such there have been a number of studies examining the best way in which to utilise them \cite{yoash18, young22}. Optimal sensor placement has also been based upon the surveillance volume coverage approach, resulting in cookie-cutter models for detection performance \cite{fewell11, craparo}. Analysis of the design of multistatic sonobuoy fields for regional surveillance has also been considered \cite{ozols11}. These analyses either suppose an active sonar or adopt a maximal-detection range approximation to assess performance.

Optimisation of passive sonar systems has been of interest for many years, with examples being \cite{pasupathy78} which is concerned with optimal spatial signal processing, \cite{grindon81} who examines mult-sensor source location problems, \cite{neering07} which considers target localisation and \cite{gao16} which is focused on the optimal design of broadband passive sensors. Tracking and detection of threats with passive systems have been explored in a number of studies \cite{yocum11, peng19, ken22, wang23}. The problems of passive sonar performance in littoral environments is examined in \cite{yuan06}. Passive sonar systems deployed in an autonomous uncrewed underwater system has also been investigated \cite{glegg01}.

In order to determine the best placement of passive sensors in a surveillance region it is necessary to develop a suitable model for single sensor performance prediction. This can be achieved through analysis of the passive sonar equation \cite{tiel76}. The latter provides a measure of the signal to noise ratio (SNR) in a logarithmic scale, and is referred to as the signal excess (SE). This is a function of sonar characteristics as well as the operational environment \cite{urick}. Two components of the SE are stochastic in nature, namely the source and noise level.
The former is a measure of the signal received at the hydrophone while the latter measures localised interference due to the environment. Such noise may be attributed to shipping as well as that generated by biological entities \cite{wang16}. The approach taken in this paper is to apply known source levels to assess performance. Such source levels of interest, at various frequencies, are available in a number of references \cite{urick, coppens76}.

The noise level can be modelled by statistics commonly used to model background noise in sonar processes \cite{ainslie10}. A Gaussian baseband model is often assumed, and depending on the signal processing either an exponential or Rayleigh distribution is then used for noise. Mean noise levels are also documented relative to particular noise sources \cite{miasnikov94}. Under either of these models it is then possible to develop a statistical representation for the SE and consequently an expression for a single sensor probability of detection. Extending this to a series of passive sensors then requires a sensor fusion protocol to be specified \cite{li21}. Due to the fact that passive sensors have limited range it is useful to develop a fusion model based upon 1 out of $N$ detection \cite{ainslie10}, where $N$ is the number of sensors. This is equivalent to declaring a target present in a surveillance region if at least one sensor registers a threat.
Since detection will be a function of the target location, it is necessary to develop a way to model this uncertainty. Hence target location is modelled through a uniform distribution over the surveillance region, which is equivalent to assuming a Bayesian non-informative prior for this location. This then permits the specification of the probability of system-level detection as a function of sensor locations. By defining a suitable objective function it will be demonstrated how the optimal sensor locations may be determined.

A significant complexity in the study of sonar performance analysis is modelling the transmission loss (TL) , which measures the signal attenuation at the sensor. TL is influenced by a number of factors including sound speed profiles, bathymetry, spreading, absorption and bottom loss \cite{urick}. In order to undertake realistic sonar performance prediction an environmental model is assumed and then applied to a computational acoustics package to calculate the TL \cite{yuan06}. In this paper it will be assumed that the sound speed remains constant and that a spherical spreading TL model is suitable, as described in \cite{tyler92, ainslie10}. These TL models include factors accounting for frequency-dependent absorption. However it will become clear that an emprically determined TL may be applied to the models and algorithms developed in this paper.

\section{Fundamental Mathematical Model}
\label{sec:model}
Suppose that a region of interest has a required surveillance volume $\cal V$ and it is desired to deploy a series of $N$ passive sensors to monitor for threats. It is assumed in this analysis that the sensor locations are fixed. Denote by ${\cal S}_1, {\cal S}_2, \ldots, {\cal S}_N$ the coordinates within $\cal V$ of each such sensor. Assume that there is a sensor fusion protocol in place such that $\pd({\cal S}_1, {\cal S}_2, \ldots, {\cal S}_N)$ is the probability of detection of a threat, at a given time instant, by the sensor network. Then the problem under consideration is to maximise this detection probability over all possible sensor locations. This probability of detection may also be viewed as an objective function for the purposes of optimisation. The following will demonstrate how sensor locations can be determined in the 2-dimensional setting; the extension to 3-dimensions proceeds along similar lines but increases the computational complexity.

Towards this aim, suppose initially that the surveillance region is a square of length $R$, defined through the Cartesian product $[0, R] \times [0, R]$, for some known $R > 0$.
Assume that sensor $j$ is located at coordinates ${\cal S}_j = (Sx_j, Sy_j)$ for each $j \in \{1, 2, \ldots, N\}$. It is postulated that there is a threat located in the surveillance region, whose location and bearing is unknown. Since this study will examine detection snapshots it is supposed that at a given time the threat is located at coordinates $(x_0, y_0)$. From a modelling perspective one may assume that this threat is equally likely located anywhere within the surveillance region. Hence it will be supposed that each of these threat location coordinates is uniformly distributed on an interval of length $R$. Equivalently, in the language of Bayesian inference,  it is being assumed that the location is modelled through a non-informative uniform prior.
Assuming that these coordinates are independent the joint density is given by
\begin{equation}
f_{(x_0, y_0)}(x, y) = \frac{1}{R^2}, \label{threatloc}
\end{equation}
provided $0 < x_0, y_0 < R$ and is zero otherwise. 

Denote by $\P_j(x, y)$ the probability that sensor $j$ detects the threat given the threat is located at coordinates $(x, y)$. It is assumed that the sensors operate independently. From a sensor fusion perspective, there are a number of ways in which threat detection may be determined. Since passive sensors will be range-limited it is suitable to utilise a sensor fusion model which is based upon 1 out of $N$ detection, or declaring a target present if at least one sensor detects a threat. As will become apparent, this is necessary from a surveillance region coverage perspective, since it may be difficult for more than one sensor to record a detection if sensor detection ranges are limited. Under such a target detection protocol, 
\begin{eqnarray}
\pd({\cal S}_1, {\cal S}_2, \ldots, {\cal S}_N) &=&   \prob(\mbox{At least one sensor detects it})\nonumber\\
&=& 1 - \prob(\mbox{No sensor detects the threat})\nonumber\\
&=& 1 - \int_0^R \int_0^R \prob(\mbox{No sensor detects the threat given threat located at }(x, y)) f_{(x_0, y_0)}(x, y) dx dy, \nonumber \\
\label{det1}
\end{eqnarray}
where $\prob$ denotes probability and conditional probability has been applied in \eqref{det1}. By applying the density \eqref{threatloc} and $ \P_j(x, y)$ to \eqref{det1} it follows that
\begin{equation}
\pd({\cal S}_1, {\cal S}_2, \ldots, {\cal S}_N)  = 1 - \frac{1}{R^2} \int_0^R \int_0^R \prod_{j=1}^N \left[1 - \P_j(x, y)\right] dx dy, \label{det2}
\end{equation}
where independent detections has been utilised to introduce the product in the above expression.
Consequently, for a given configuration of the $N$ sensors within the surveillance region, this expression provides a quantitiative measure of the likelihood of detection of the threat whose position is unknown. 

\begin{figure}[h]
\centering
\includegraphics[scale =0.8]{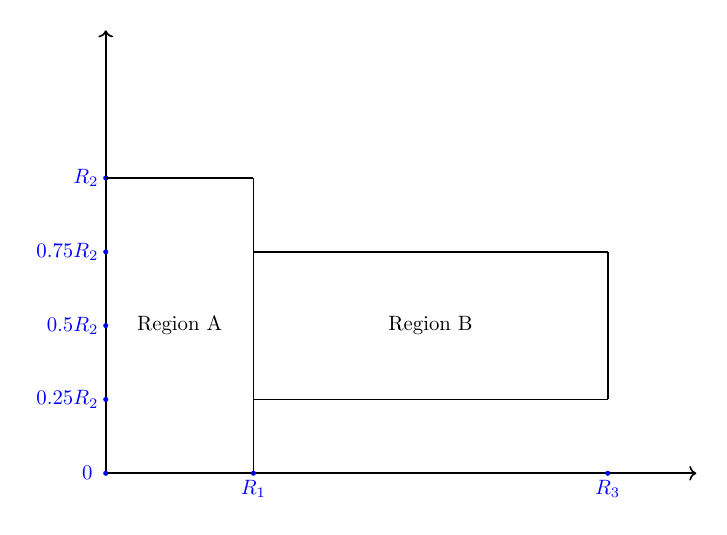}
\caption{An example of a hybrid region used for model development.}
\label{figRegion}
\end{figure}

One may extend this analysis to include the problem of sensor placement in hybrid regions, such as that illustrated in Figure \ref{figRegion}. This has been motivated by the problem of sensor placement in a chokepoint. As an example, this idealised scenario may function as a guide on how to place sensors in the Strait of Gibraltar, where Region A represents a surveillance area of the Atlantic Ocean and Region B represents the Strait. Separation of the surveillance region into two such components then allows separate models to be introduced to capture differing noise levels and environmental factors, some of which may contribute to the difficulty in detecting an undersea threat \cite{renee24}. It is also worth pointing out the fact that in such regions of increased acoustic activity there is much concern of the impact of underwater sound on marine life. Hence in such scenarios passive sensors provide an environmentally safer means of surveillance \cite{folegot12}.

To develop a modelling framework for scenarios such as that in Figure \ref{figRegion}, suppose that there is a single threat in this hybrid region. Region A is bounded by points $(0, 0), (R_1, 0), (R_1, R_2)$ and $(0, R_2)$. In the case of Region B, the boundary points are $(R_1, 0.25R_2)$, $(R_3, 0.25 R_1)$, $(R_3, 0.75 R_2)$ and $(R_1, 0.75 R_2)$.  Then in absence of target location information, the probability that it lies in either region becomes one of an area comparison. Hence
\begin{equation}
\prob(\mbox{Threat in Region A}) = \frac{R_1 R_2}{R_1 R_2 + 0.5R_2(R_3 - R_1)}, \label{probA}
\end{equation}
while the probability that it is Region B is the complement of \eqref{probA}. Within each region \eqref{det2} may be adapted as follows. 
Denote by $\{{\cal S}^1_j, j \in \{1, 2, \ldots, N_1\}\}$ the sensor location coordinates in Region A, where $N_1$ is the number of such sensors. Similarly, let 
 $\{{\cal S}^2_j, j \in \{1, 2, \ldots, N_2\}\}$ be the corresponding sensor locations in Region B, with $N_2$ the number of sensors. Then 
for the first region,
\begin{equation}
\pd({\cal S}_1^1, {\cal S}_2^1, \ldots, {\cal S}_{N_1}^1| \mbox{Threat in Region A}) 
  = 1 - \frac{1}{R_1 R_2} \int_0^{R_1} \int_0^{R_2} \prod_{j=1}^{N_1} \left[1 - \P_j^1(x, y)\right] dy dx, \label{det3}
\end{equation}
while for the second region
\begin{equation}
\pd({\cal S}_1^2, {\cal S}_2^2, \ldots, {\cal S}_{N_2}^2| \mbox{Threat in Region B})   = 1 - \frac{1}{0.5R_2(R_3 - R_1)} \int_{R_1}^{R_3} \int_{0.25R_2}^{0.75 R_2} \prod_{j=1}^{N_2} \left[1 - \P_j^2(x, y)\right] dy dx, \label{det4}
\end{equation}
where $ \P_j^i(x, y)$ is the single sensor probability of detection in each respective region ($i \in \{1, 2\}$). Expressions \eqref{det3} and \eqref{det4} are modelling the threat location within the corresponding region using a uniform distribution over the surveillance area.
Observing that by Bayes' Theorem
\begin{eqnarray}
\prob(\mbox{Threat Detected in Hybrid Region}) &=& \pd({\cal S}_1^1, {\cal S}_2^1, \ldots, {\cal S}_{N_1}^1| \mbox{Threat in Region A})  \prob(\mbox{Threat in Region A})\nonumber\\
  &+& \pd({\cal S}_1^2, {\cal S}_2^2, \ldots, {\cal S}_{N_2}^2| \mbox{Threat in Region B}) \prob(\mbox{Threat in Region B}) \label{det5}
\end{eqnarray}
and thus one may apply \eqref{probA}, \eqref{det3} and \eqref{det4} directly to \eqref{det5} to produce the probability of threat detection in the hybrid region, as a function of the two sets of sensors. It is useful to observe that since \eqref{det5} is the sum of two non-negative components that in order to maximise the probability of detection over the hybrid region it is sufficient to maximise sensor placement within each subregion.

What is required for evaluation of \eqref{det2} and \eqref{det5}  is an expression for the probability of detection of the threat from sensor $j$, namely $\P_j(x, y)$ or $\P_j^i(x, y)$. In order to do this it is necessary to examine the well-known passive sonar equation. This will be the focus of the following section.

\section{Single Sensor Probability of Detection}
\label{sec:spd}
Since the focus in this paper is on the passive sonar detection of stealthy threats it will be assumed that the hydrophones are broadband passive sensors. Hence it will be necessary to formulate some components in the passive sonar equation for this context \cite{ainslie10}. 
The passive sonar equation provides a measure of the SE, in terms of a number of influential sonar parameters, which provides a decibel (dB) measure of the SNR received at a particular sensor \cite{coppens76, biediger10}. In the case where the SE exceeds zero the presence of a threat will be confirmed. The passive sonar equation is
\begin{equation}
SE = SL - TL - NL + DI - DT, \label{passivesonar}
\end{equation}
where SL is the source level, or equivalently the dB strength of a signal received at the sensor, TL is the transmission loss, which is a measure of the attenuation of the signal in the maritime environment. NL is the noise level which captures acoustic interference from the environment, either due to natural phenomena or naval vessels. DI is the directivity index, which provides a measure of spatial directivity of the receiving hydrophone, also providing a quantification of the amount of noise reduction the sensor provides. Finally, DT is the detection threshold, which provides a measure of the required SNR to achieve a probability of detection ($\pd$) for prescribed probability of false alarm ($\pfa$). This equation is analysed in considerable detail in sonar textbooks \cite{urick, ainslie10} but more compact analyses can be found in \cite{tyler92}.
Some of the terms in \eqref{passivesonar} will be known {\em a priori}, such as the DI and the DT. The former may be specified based upon the deployed hydrophone array characteristics, while the latter will be determined by operational requirements. In particular, a useful guide on the DT is \cite{dawe} who explains suitable selections for it for passive broadband sonar. Both the TL and NL will be functions of the operating environment, with TL being a range-dependent function. The SL and NL are statistical in nature; throughout the following a suitable model will be formulated for the NL and an average over frequency values will be used to estimate the SL for problems of interest.

Based upon the discussion in \cite{coppens76} one may decompose both the source and noise level to the sum of the spectrum levels and the sum of the system bandwidth in dB. The latter terms then cancels out and one can instead consider the passive sonar equation
\begin{equation}
SE = SSL - TL - NSL + DI - DT, \label{passivesonar2}
\end{equation}
where SSL is the source spectrum level and NSL is the noise spectrum level. In the following it is assumed that the background statistics are Rayleigh distributed and that detection is based upon amplitude statistic considerations. Under this assumption the ambient noise model is also based upon Rayleigh statistics \cite{dawe}. 

Since the probability of detection is measured by the event that the SE exceeds zero, it follows that
\begin{eqnarray}
\prob(\mbox{Detection Made}) &=& \prob(SE > 0) \nonumber\\
&=& \prob(NSL < SSL - TL - DT + DI) \nonumber \\
&=& F_{NSL}( SSL - TL - DT + DI), \label{detA}
\end{eqnarray}
where $F_{NSL}$ is the distribution function of the NSL. Hence a model for the NSL is required.

Suppose that at the baseband level the NSL has a frequency-dependent Rayleigh distribution; hence suppose that $NSL(f)$ is the NSL at the narrowband frequency $f$. Then 
we write $NSL(f) \dist \ray (\sigma_f)$ to indicate that $NSL(f)$ has a Rayleigh distribution with distribution function
\begin{equation}
\prob(NSL(f) \leq t) = 1 - e^{-\frac{t^2}{2\sigma_f^2}}, \label{raycdf}
\end{equation}
where $t \geq 0$ and $\sigma_f$ is the associated distributional parameter. Observe that if one defines $Y = (\sqrt{2} \sigma_f)^{-1} NSL(f)$ then it can be shown that $Y \dist \ray (1/\sqrt{2})$ and so $NSL(f) = \sqrt{2} \sigma_f Y$.

Assume that the noise spectrum has center frequency $f_m$ Hz and bandwidth $B$ Hz. Then based upon the discussion in \cite{ainslie10} the broadband NSL is given by
\begin{equation}
NSL = 10 \log_{10} NSL^*, \label{nsl1}
\end{equation}
where
\begin{equation}
NSL^* = \frac{1}{B} \int_{f_m - \frac{B}{2}}^{f_m + \frac{B}{2}} NSL(f) df, \label{nsl2}
\end{equation}
which is the frequency-dependent NSL averaged over the frequency band. By noting that 
\begin{equation}
\mean NSL(f) = \sqrt{2} \sigma_f \mean(Y) = \sqrt{\frac{\pi}{2}} \sigma_f, \label{nsl3}
\end{equation}
where $\mean$ is the statistical mean of the associated random variable, 
it follows by applying \eqref{nsl3} to \eqref{nsl2} that
\begin{equation}
NSL^* = \frac{2}{\sqrt{\pi} B} \left(\int_{f_m - \frac{B}{2}}^{f_m + \frac{B}{2}} \mean NSL(f) df \right)Y. \label{nsl4}
\end{equation}
Consequently, an application of \eqref{nsl4} to \eqref{nsl1} results in 
\begin{equation}
NSL = 10 \log_{10}\left[ \frac{2}{\sqrt{\pi} B} \int_{f_m - \frac{B}{2}}^{f_m + \frac{B}{2}} \mean NSL(f) df \right] + 10 \log_{10} Y.
\label{nsl5}
\end{equation}
The first component in \eqref{nsl5} is deterministic and constant, while the second is stochastic. Note that
\begin{equation}
\overline{NSL} := \frac{1}{B} \int_{f_m - \frac{B}{2}}^{f_m + \frac{B}{2}} \mean NSL(f) df \label{nslmean}
\end{equation}
is the average over the frequency band of the average NSL at each frequency $f$. This may be estimated based upon data as in \cite{coppens76}.
If $\gamma$ is defined to be the first term in \eqref{nsl5} then one can show with an application of \eqref{nsl5} that \eqref{detA} reduces to 
\begin{equation}
\prob(\mbox{Detection Made}) = 1 - e^{-10^{0.2[SSL - TL - DT + DI - \gamma]}}, \label{nsl6}
\end{equation}
where the Rayleigh distribution function of the form \eqref{raycdf} with $\sigma_f^2 = 0.5$ has been utilised.

Hence this expression provides one with the single-sensor probability of detection, which may be applied for $\P_j(x, y)$ defined in the previous section. What remains is to specify the remaining non-stochastic sonar terms in \eqref{nsl6}. 

Firstly consider the SSL. In the broadband case it is given by
\begin{equation}
SSL = 10 \log_{10}(SSL^*), 
\label{ssl1}
\end{equation}
where 
\begin{equation}
SSL^{*} = \frac{1}{B} \int_{f_m - \frac{B}{2}}^{f_m + \frac{B}{2}} SSL(f) df, \label{ssl2}
\end{equation}
which is the frequency-dependent SSL averaged over the frequency band\cite{ainslie10}. 
For the applications to follow this will be based upon measured levels for targets of interest.

Next the TL is examined. In the broadband case this is also averaged over frequency, and under the assumption of linear variation of attenuation with frequency a simplified expression for broadband TL is described in \cite{ainslie10}. In particular when the absorption coefficient is given by
\begin{equation}
\alpha(f) = \alpha_m + \alpha_f (f - f_m), \label{alphadef}
\end{equation}
where $\alpha_m$ and $\alpha_f$ are constants, in units of dB per kiloyard, the broadband TL is given by
\begin{equation}
TL = -10\log_{10} F_{BB} \label{tl1}
\end{equation}
where
\begin{equation}
F_{BB} = \frac{2}{r^2} e^{-2\alpha_m r} {\rm sinhc}(\alpha_f B r).
\label{tl2}
\end{equation}
The range $r$ is either in kiloyards or nautical miles, but then requires adjustment to the units of the absorption coefficient.
An application of \eqref{tl2} to \eqref{tl1} yields the expression
\begin{equation}
TL(r) = (20 \alpha_m \log_{10}e)r + 30 \log_{10}r + 10 \log_{10}(\alpha_f B) - 10 \log_{10}\left[ e^{\alpha_f B r}  - e^{-\alpha_f Br} \right],\label{tl3}
\end{equation}
which can be applied to \eqref{nsl6} and $r$ is in kiloyards. For the case where $r$ is measured in nautical miles, this becomes
\begin{equation}
TL(r) = (40 \alpha_m \log_{10}e)r + 30 \log_{10}r + 10 \log_{10}(2\alpha_f B) - 10 \log_{10}\left[ e^{2\alpha_f B r}  - e^{-2\alpha_f Br} \right], \label{tl3a}
\end{equation}
where a nautical mile has been approximated by two kiloyards.
For the puposes of this study, the result derived in \cite{thorp67} may be applied for the frequency-dependent absorption coefficient. The more refined expression in \cite{tyler92} is instead used and a linear interpolation between the limits of the broadband frequency range is employed in the numerical analysis.

When the bandwidth $B$ is small it can be shown that \eqref{tl3} reduces to the well-known expression for spherical spreading.
For any given sensor $j$ the range $r$ in \eqref{tl3} will be the Euclidean distance between the sensor located at $(Sx_j, Sy_j)$ and the target location $(x, y)$.

Since we are concerned with broadband passive sonar, based upon the analysis reported in \cite{dawe}, if we assume that the detection process is based upon amplitude statistics then the underlying noise distributions can be modelled as Rayleigh for small number of samples and when the SNR is expected to be small. Hence one may apply the expression
\begin{equation}
DT = 10\log_{10}\left( \frac{0.273d}{t} + 1.045 \sqrt{\frac{d}{t}}\right) \label{detThr}
\end{equation}
for the detection threshold (which is relative to a bandwidth $B$), with $t$ being the time to make a detection decision and where the detection index is
\begin{equation}
d = \left(\frac{\pi}{4-\pi}\right)\left( \sqrt{\frac{\log_e \pfa}{\log_e \pd}} - 1\right)^2, \label{dindx}
\end{equation}
for an integration factor of unity, for a desired $\pd$ and $\pfa$. Since this analysis is concerned with initial threat detection it may be supposed that $t = 1$ in \eqref{detThr}. 

The final term to consider is the DI. In the case of sensors deployed in a surveillance volume one may suppose that these are omnidirectional, so that $DI = 0$.  However, in scenarios such as that depicted in Figure \ref{fig1} it may be possible to deploy fixed arrays along sections such as in Region B. In this way these sensors will provide enhanced detection performance. Under the assumption of a unsteered linear planar array it can be shown that the DI is approximately
\begin{equation}
DI \approx 10\log_{10} (DI^*) \label{BBDI1}
\end{equation}
and for the narrowband case $DI^* = \frac{2L}{\lambda^2}$, where $L$ is the array length and $\lambda$ is the wavelength. In order to extend this to the broadband case one may average over the wavelength instead of frequency. Hence if $\lambda_1$ and $\lambda_2$ are the wavelengths corresponding to the previously assumed broadband frequency band then 
\begin{equation}
DI^* = \frac{1}{\lambda_2 - \lambda_1} \int_{\lambda_1}^{\lambda_2}  \frac{2L}{\lambda^2} d\lambda = \frac{2L}{\lambda_1 \lambda_2}, \label{BBDI2}
\end{equation}
which when applied to \eqref{BBDI1} results in
\begin{equation}
DI \approx 10\log_{10}\left(\frac{2L}{\lambda_1 \lambda_2}\right). \label{BBDI3}
\end{equation}
For the linear array to provide gains it is necessary to assume that $L >> 0.5 \lambda_1 \lambda_2$.

Expressions \eqref{tl3} (or \eqref{tl3a}) and  \eqref{detThr} may be applied to \eqref{nsl6}, with either $DI$ set to zero or defined through \eqref{BBDI3}, and then utilised in \eqref{det2} and \eqref{det5} to assess the performance of a given set of passive sensors, provided estimates of relevant parameters are applied. Hence the next section will provide a series of examples to illustrate this. Additionally, it will be shown how their placement may be determined optimally within regions of interest through the analysis undertaken.

\section{Applications}
\label{sec:app}

\subsection{Study Parameters}
Throughout the following all ranges are measured in nautical miles (nmi) and the speed of sound is taken to be constant $c = 1500$ metres per second.
In order to provide tangible examples of performance it is necessary to adopt some model parameterisations. The purpose of this subsection is to provide clarification of these and to provide some justification for choices adopted. Throughout the broadband centre frequency is taken to be $f_m = 100$ Hz with a $B = 30$ Hz bandwidth. With this choice one can obtain estimates of  SSL and NSL from the literature. For the NSL, Table 6.5 in \cite{bjorno17} allows one to obtain estimates of the mean NSL across the adopted broadband frequency band. A linear interpolation was applied to estimate \eqref{nslmean} resulting in the choice of $\overline{NSL} = 68.5$ dB. Both \cite{urick} and \cite{coppens76} provide estimates of the SSL for conventional submarines on electric drive operating at periscope depth as a function of frequency and speed. A linear interpolation was applied for the case of a speed of 4 kts and the resulting estimate of \eqref{ssl1} was 133 dB. 

To provide estimates of the parameters $\alpha_m$ and $\alpha_f$ in the absorption coefficient given by \eqref{alphadef} a linear interpolation has also been applied over the frequency band.
By applying the expression for the absorption coefficient in \cite{tyler92}, given by equation (9) in the latter but adjusted to dB per nmi, one extracts $\alpha_f = 0.20347$ and $\alpha_m = 31.2218$.

In the case of the DT \eqref{detThr} the probability of detection was set to 0.5 with a probability of false alarm of $10^{-4}$. Hence $DT = 10.8917$ is applied uniformly in all cases.

When the DI is given by \eqref{BBDI3}, the wavelengths corresponding to the frequency band are $\lambda_1 = 13.04$ and $\lambda_2 = 17.65$ metres and so 
$DI = 10\log_{10}(L) - 20.61$ dB with the requirement that $L > 115.08$ in metres.

\subsection{Single Sensor Performance: Omnidirectional Case}
\begin{figure}[h]
\centering
\includegraphics[scale =0.5]{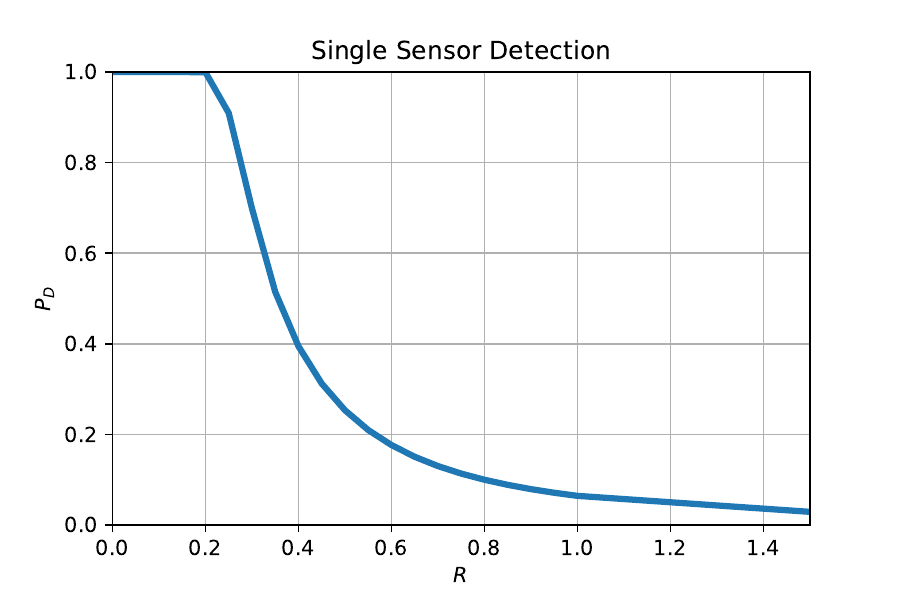}
\caption{Single omnidirectional sensor performance as a function of the surveillance boundary parameter $R$.}
\label{fig1}
\end{figure}

In the first instance, the performance of a single omnidirectional receiver is examined, where it is located in the centre of the surveillance region. The latter is taken to be a square of length $R$, and Figure \ref{fig1} plots \eqref{nsl6} as a function of $R$ under the parameters outlined above. What this figure shows is that detecton is certain for targets located within 0.2 nmi. At double this range, the probability of detection reduces to 0.4. This demonstrates the limits of passive sensing through a single omnidirectional receiver. This also indicates how many such sensors would be required to cover a given region. For a given $R$, to provide coverage of the square region of length $R$, one requires $R^2/0.2^2$ sensors. In the case where $R = 0.4$ nmi the number of sensors would be 4, where these are placed in the centre of four subregions of the surveillance region. When $R$ is increased to 1 nmi, a total of 25
sensors would be required.

There is another important consequence of this example. It illustrates that a surveillance area may be sufficiently covered by embedding sensors within squares. In the usual cookie-cutter appproach detection circles are investigated, which either results in overlapping sensor regions or insufficiently covered regions. This issue is rectified through this new approach, and the next subsection will illustrate it further.

\subsection{Multiple Omnidirectional Sensor Examples}
\label{subsecC}
\begin{figure}[h]
\centering
\includegraphics[scale =0.5]{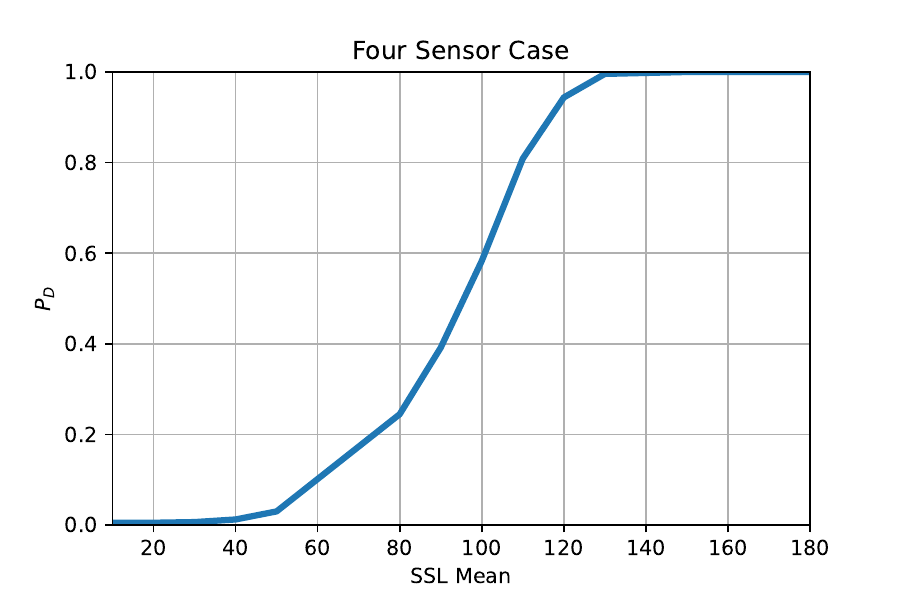}
\caption{Probability of detection as a function of the average SSL, for four sensors in a 0.4 nmi square.}
\label{fig2}
\end{figure}

Suppose that the surveillance region is a square of length $R = 0.4$ nmi, and so from the previous discussions one requires four omnidirectional sensors to maximise coverage.
Suppose that these are deployed at coordinates $(0.1, 0.1), (0.3, 0.1), (0.1, 0.3)$ and $(0.3, 0.3)$, which are the centres of four squares embedded in the region. Figure \ref{fig2}
plots the probability of detection as a function of SSL mean but for the fixed average noise level stipulated above. At the mean SSL level of 133 dB the probability of detection is almost unity. Below this average SSL the probability of detection is significantly smaller. The figure illustrates the difficulty of sensing a target with small average SSL.

\begin{figure}[h]
\centering
\includegraphics[scale =0.5]{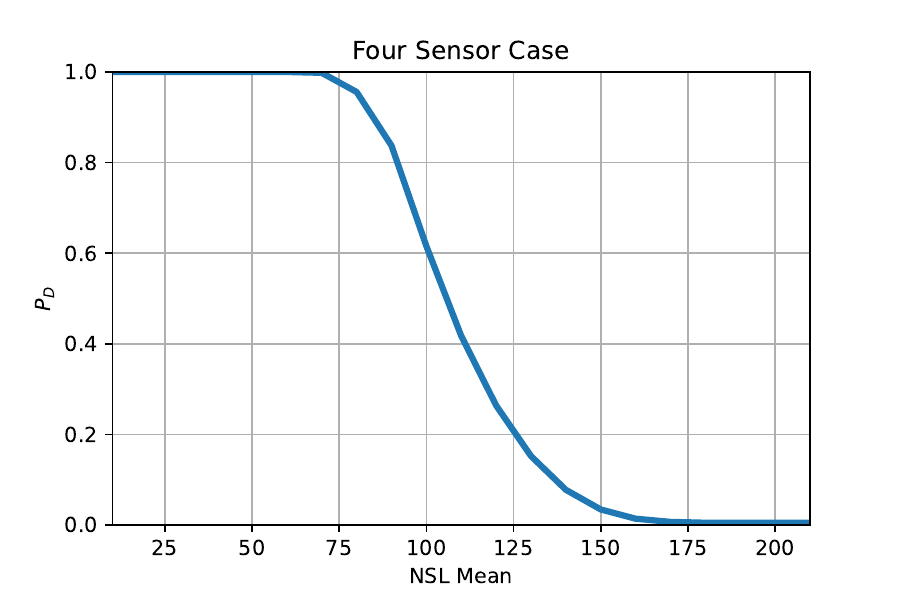}
\caption{An alterative perspective of Figure \ref{fig2}. Here the detection probability is a function of the average NSL.}
\label{fig3}
\end{figure}

To provide another perspective on this situation, Figure \ref{fig3} plots the probability of detection as a function of the average NSL, where the average SSL is set to 133 dB. The figure implies that provided the average NSL is smaller than 75 dB target detection is certain.

\begin{figure}[h]
\centering
\includegraphics[scale =0.5]{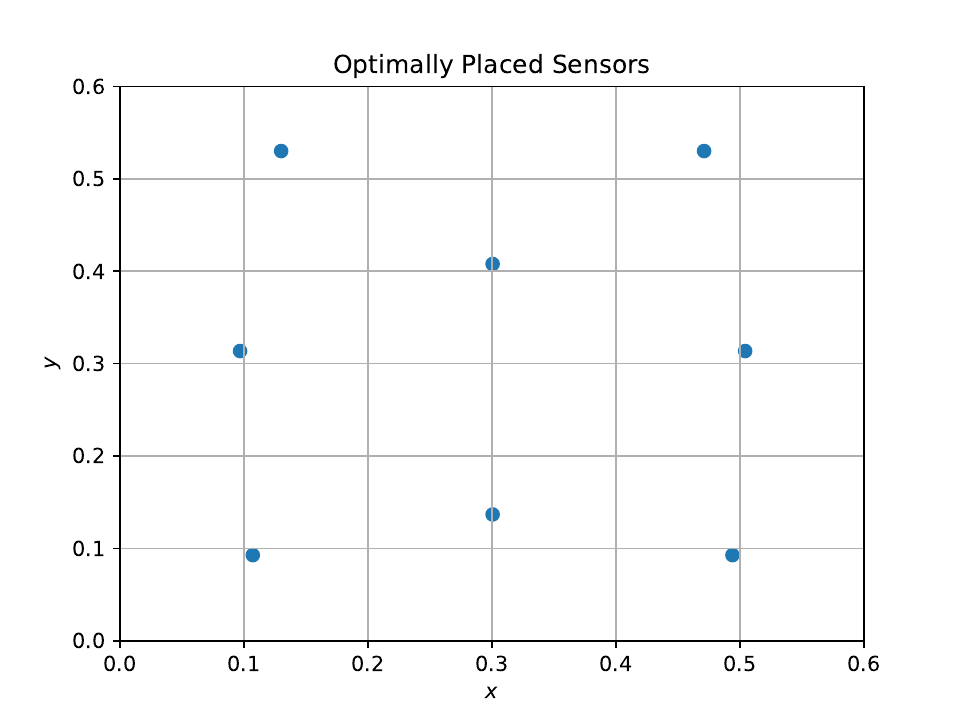}
\caption{Optimal sensor placement, as determined through optimisation of an objective function in Python, for the case where $R = 0.6$ nmi.}
\label{figsensorplacement}
\end{figure}

In some square surveillance regions it may not be possible to embed a series of squares with sensors located at the centre of each. An example is the case where $R = 0.6$ nmi, where the number of required sensors is 9. In such cases it is useful to utilise an optimisation algorithm to facilitate determination of sensor locations. By defining the probability of detection as an objective function of the sensor locations, a differential evolution algorithm was applied in Python to determine the optimal sensor locations. In this situation the complement of the probability of detection was defined as the objective function, and the algorithm searched for a local minimum over sensor locations. For this analysis it was found that 8 sensors would be sufficient to provide coverage of the area under consideration. The algorithm yielded the sensor location vector whose coordinates are provided in Table \ref{table1}, and this resulted in a probability of detection of 0.98. An illustration of the optimal sensor locations is provided in Figure \ref{figsensorplacement}. It is interesting to observe the symmetry around the $x = 0.3$ axis.

\begin{table}
\begin{center}
\begin{tabular}{||cccc||}
\hline &&& \\[-2ex]
(0.49390946, 0.09272115) & (0.50415978, 0.31359301) & (0.12996332, 0.53003436) & (0.09682469, 0.31359738) \\
 (0.47103453, 0.5300298)  & (0.30049256, 0.40783295) & (0.30049426, 0.13672337)  & (0.30049426, 0.13672337) \\
\hline
\end{tabular}
\end{center}
\caption{Optimal sensor coordinates for the region with $R = 0.6$ nmi.}
\label{table1}
\end{table}

\subsection{Single Sensor Performance: Fixed Array Case}
\label{subsecD}
\begin{figure}[h]
\centering
\includegraphics[scale =0.8]{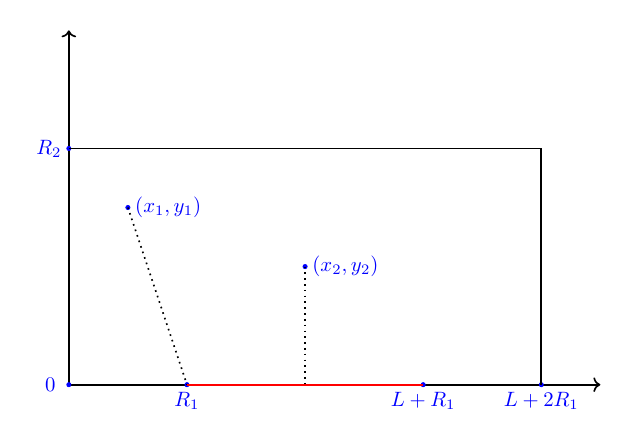}
\caption{Surveillance region for the deployment of a fixed linear array. The array of length $L$ is located along the $x$-axis, indicated in red. The distance between a threat and the array is the least distance between these two points, as illustrated with two potential target locations.}
\label{figRegion2}
\end{figure}

A fixed unsteered planar array may be placed along the boundary of a region to provide enhanced detection performance. Consider a region as illustrated in Figure \ref{figRegion2}, where an array of length $L$ is located as shown, which runs between coordinates $(R_1, 0)$ and $(R_1 + L, 0)$. The region is a rectangle with dimensions $(L+2R_1) \times R_2$, so 
that one may assume a target is uniformly distributed in this area. In order to apply \eqref{tl3a} one must specify the distance $r$ between the sensor and the target location, for application in \eqref{nsl6}. Since the array is of length $L$, for any given target location $(x, y)$ one must select $r$ to be the minimum distance between the point $(z+R_1, 0)$ and $(x, y)$, where $0 \leq z \leq L$. Figure \ref{figRegion2} provides an illustration of the selection of minimum distance. Thus it is being assumed that a wavefront propagating from the target will impinge on the receiver located closest to the point source.

\begin{figure}[h]
\centering
\includegraphics[scale =0.5]{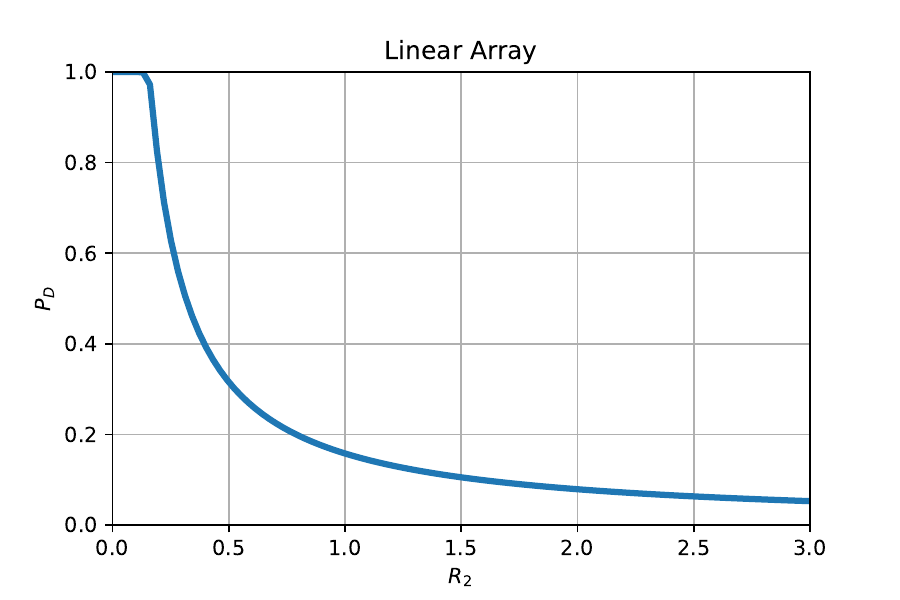}
\caption{Probability of detection for a region such as that in Figure \ref{figRegion2} with $R_1 = 0.1$ nmi and array length $L = 0.8$ nmi.}
\label{figlinearray}
\end{figure}

As an example, Figure \ref{figlinearray} plots the probability of detection as a function of $R_2$ for the case where $R_1 = 0.1$ nmi and $L = 0.8$. Hence the rectangle in Figure \ref{figRegion2} is 1 nmi along the $x$-axis, with $R_2$ varying from 0 to 2 nmi along the $y$-axis. This result shows that for this linear array the probability of detection will be at least 0.5 for $R_2$ less than 0.25 nmi. This figure also implies that such a linear array will provide reasonable coverage for narrow regions of dimensions 0.2 $\times$ 1 nmi, where the probability of threat detection is almost certain. For wider regions, one may improve this if it is possible to deploy a second linear array parallel to the first. Based upon these considerations, such a deployment should provide sufficient coverage for a region of dimensions 0.4 $\times$ 1 nmi.

\subsection{Hybrid Region Example}
\begin{figure}[h]
\centering
\includegraphics[scale =0.5]{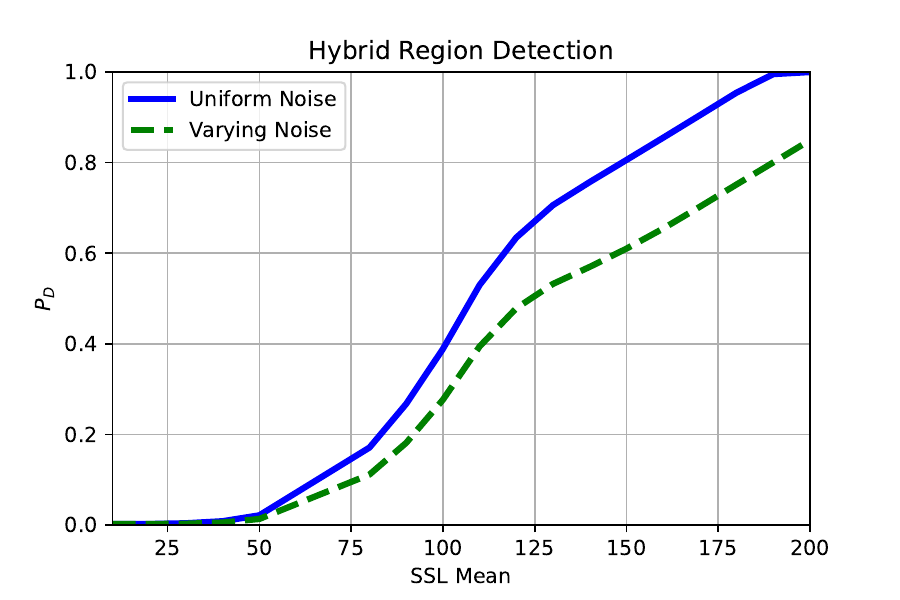}
\caption{Probability of detection for a hybrid region as in Figure \ref{figRegion} where $R_1 = R_2 = 0.4$ and $R_3 = 1.4$. Uniform noise refers to the case where the average noise level is 68.5 dB, while varying noise refers to the case where the average noise level in Region B is 109.6 dB.}
\label{fighybrid}
\end{figure}

Consider the case of a hybrid region as in Figure \ref{figRegion} where Region A has $R_1 = R_2 = 0.4$, Region B is 1 nmi in length so that $R_3 = 1.4$ and so the width of Region B is 0.2 nmi. Hence this is constructed from the examples considered previously. Therefore the placement of sensors in Region A will be as specified in Subsection \ref{subsecC}, while a linear array of length 0.8 nmi is positioned along a boundary in Region B as in Subsection \ref{subsecD}. Two cases will be considered. The first is where the average noise level is uniform across both regions (68.5 dB), while for the second case the noise level in Region B is assumed to be 60\% greater (109.6 dB).
Figure \ref{fighybrid} plots the probability of detection \eqref{det5} as a function of the SSL mean for both these cases. The figure illustrates the difficulty in detecting small SSL targets and the impact of increased noise on detection performance. 

\section{Conclusions}
The purpose of this paper has been to demonstrate how passive sonar placement may be determined through the probability of threat detection.
In order to achieve this a new expression has been produced for the single sensor probability of detection from a passive broadband sonar. Examples were then used to 
illustrate how the results may be applied to determine sensor placement, both in simple and more complex regions. For simple square regions the approach indicated that one may determine detection coverage by embedding a series of squares within the region, with the sensor placed at the centre of each such square. The number of required sensors could then be determined. In cases where it is not possible to partition the area into uniform squares one could instead apply a numerical optimisation algorithm to determine sensor placement. 
This approach was then extended to include unsteered linear arrays and applied to hybrid regions such as those arising near chokepoints.

In futher work it would be useful to explore the application of these
results to non-uniform surveillance regions, as well as the extension to the 3-dimensional setting.



\end{document}